\documentclass{PoS}
\usepackage{amsmath}
\usepackage{graphicx}
\usepackage{cite}
\title{B-anomalies related to leptons and lepton flavour universality violation}

\ShortTitle{B-anomalies}

\author{\speaker{Andreas Crivellin}\thanks{I thank the organizers for the invitation and the opportunity to present these results. This work is supported by an Ambizione grant of the Swiss National Science Foundation.}\\
        Paul Scherrer Institut\\
				CH-5232 Villigen PSI\\
				Switzerland\\
        E-mail: \email{andreas.crivellin@cern.ch}}

\abstract{Several experiments observed deviations from the Standard Model (SM) in the flavour sector: LHCb found a $4-5\,\sigma$ discrepancy compared to the SM in $b\to s\mu^+\mu^-$ transitions (recently supported by an Belle analysis) and CMS reported a non-zero measurement of $h\to\mu\tau$ with a significance of $2.4\,\sigma$. Furthermore, BELLE, BABAR and LHCb founds hints for the violation of flavour universality in $B\to D^{(*)}\tau\nu$. In addition, there is the long-standing discrepancy in the anomalous magnetic moment of the muon. Interestingly, all these anomalies are related to muons and taus, while the corresponding electron channels seem to be SM like. This suggests that these deviations from the SM might be correlated and we briefly review some selected models providing simultaneous explanations.}

\FullConference{16th International Conference on B-Physics at Frontier Machines\\		2-6 May 2016\\
		Marseille, France}

\begin{document}

\section{Introduction}
\label{intro}

The discovery the Higgs at the LHC provided the final ingredient of the SM. While no direct evidence for physics beyond the SM was found during the first LHC run, there are some interesting indirect hints for new physics (NP) in the flavor sector, mainly in semileptonic decays of $B$-mesons, the SM-forbidden decay $h\to\mu\tau$ of the Higgs boson and the long-lasting discrepancy in the anomalous magnetic moment (AMM) of the muon\footnote{We do not discuss the anomaly in $\epsilon^\prime/\epsilon$~\cite{Buras:2015xba,Buras:2015yba} here for which possible solutions include $Z'$ bosons~\cite{Buras:2015kwd,Buras:2016dxz} or the MSSM~\cite{Kitahara:2016otd}.}.

{\bf\boldmath $b\to s\ell^+\ell^-$:}
Deviations from the SM found by LHCb~\cite{LHCb:2015dla} in the decay $B\to K^* \mu^+\mu^-$ arise mainly in an angular observable called $P_5^\prime$~\cite{Descotes-Genon:2013vna}, with a significance of $2$--$3\sigma$ depending on assumptions made for the hadronic uncertainties~\cite{Descotes-Genon:2014uoa,Altmannshofer:2014rta,Jager:2014rwa}. This measurement recently received support from a (less precise) BELLE measurement~\cite{Abdesselam:2016llu}. In the decay $B_s\to\phi\mu^+\mu^-$, LHCb also uncovered~\cite{Aaij:2015esa} deviations compared to the SM prediction from lattice QCD~\cite{Horgan:2013pva,Horgan:2015vla} of $3.5\sigma$ significance~\cite{Altmannshofer:2014rta}. LHCb has further observed lepton flavor universality violation (LFUV) in $B\to K\ell^+\ell^-$ decays~\cite{Aaij:2014ora} across the dilepton invariant-mass-squared range $1\,{\rm GeV}^2<m_{\ell\ell}^2<6\,{\rm GeV}^2$.  Here, the measured ratio branching fraction ratio 
$
R(K)=\frac{{\rm Br}[B\to K \mu^+\mu^-]}{{\rm Br}[B\to K e^+e^-]}
$
disagrees with the theoretically clean SM prediction by $2.6\sigma$. Combining these observables with other $b\to s$ transitions, it is found that NP is preferred over the SM by $4$--$5\sigma$~\cite{Altmannshofer:2015sma,Descotes-Genon:2015uva,Hurth:2016fbr}. Interestingly, assuming NP in muons only but not in electrons gives a better fit than assuming lepton flavour universality.

{\bf\boldmath $B\to D^{(*)}\tau\nu_\tau$:} Hints for lepton flavour universality violation (LFUV) in these modes were observed first by the BaBar collaboration~\cite{Lees:2012xj} in 2012. These measurements have been confirmed by BELLE~\cite{Huschle:2015rga,Abdesselam:2016cgx} and also LHCb has remeasured $B\to D^{*}\tau\nu_\tau$~\cite{Aaij:2015yra}. For the ratio ${R}(X)\equiv {\rm Br}[B\to X \tau \nu_\tau]/{\rm Br}[B\to X \ell \nu_\ell]$, the current HFAG average~\cite{Amhis:2014hma} of these measurements is
$
R(D)_\text{exp}=\,0.397\pm0.040\pm0.028  \,,\;\; 
R(D^*)_\text{exp}=\,0.316\pm0.016\pm0.010 \,.
$
Comparing these results to the SM predictions~\cite{Fajfer:2012vx} $R_\text{SM}(D)=0.297\pm0.017$ and $R_\text{SM}(D^*)=0.252\pm0.003$, there is a combined discrepancy of $4.0\sigma$~\cite{Amhis:2014hma}. 

{\bf\boldmath $h\to\mu\tau$:} In the Higgs sector, CMS has presented results for a search for the lepton-flavor-violating (LFV) decay mode $h\to\mu\tau$, with a preferred value~\cite{Khachatryan:2015kon}
$	{\rm Br} [h\to\mu\tau] = \left( 0.84_{-0.37}^{+0.39} \right)\%.$
This is consistent with the less precise ATLAS measurement~\cite{Aad:2015gha}, giving a combined significance for NP of $2.6\sigma$, since such a decay is forbidden in the SM. However, the first CMS run II measurement, even though not yet competitive with the run I results, does not show an excess~\cite{CMS:2016qvi}. Nonetheless, this decay mode is of considerable interest because it hints at LFV in the charged-lepton sector, whereas up to now, LFV has only been observed in the neutrino sector via oscillations. 

{\bf\boldmath $a_\mu$:} The AMM of the muon $a_\mu \equiv (g-2)_\mu/2$, provides another motivation for NP connected to muons. The experimental value of $a_\mu$ is completely dominated by the Brookhaven experiment E821~\cite{Bennett2006} and is given by $a_\mu^\mathrm{exp} = (116\,592\,091\pm54\pm33) \times 10^{-11}$, where the first error is statistical and the second systematic. The SM prediction is~\cite{Colangelo2014} $a_\mu^\mathrm{SM} = (116\,591\,855\pm59) \times 10^{-11}$, where almost the entire uncertainty is due to hadronic effects. This amounts to a discrepancy between the SM and experimental values of 
$
\Delta a_\mu = a_\mu^\mathrm{exp}-a_\mu^\mathrm{SM} = (236\pm 87)\times 10^{-11}\, ,
$
i.e.~a $2.7\sigma$ deviation\footnote{Less conservative estimates even lead to discrepancies up to $3.6\,\sigma$ in $a_\mu$}. 

\section{Explanations}

\begin{figure}[t]
\includegraphics[width=0.48\textwidth]{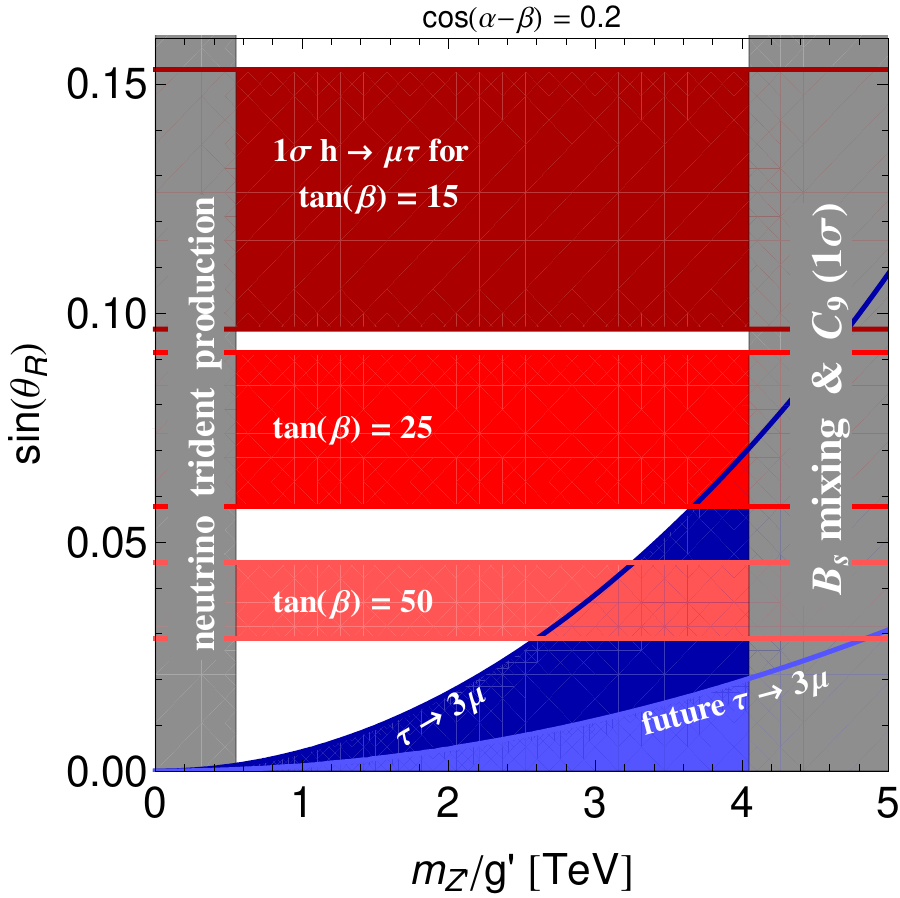}
\includegraphics[width=0.46\textwidth]{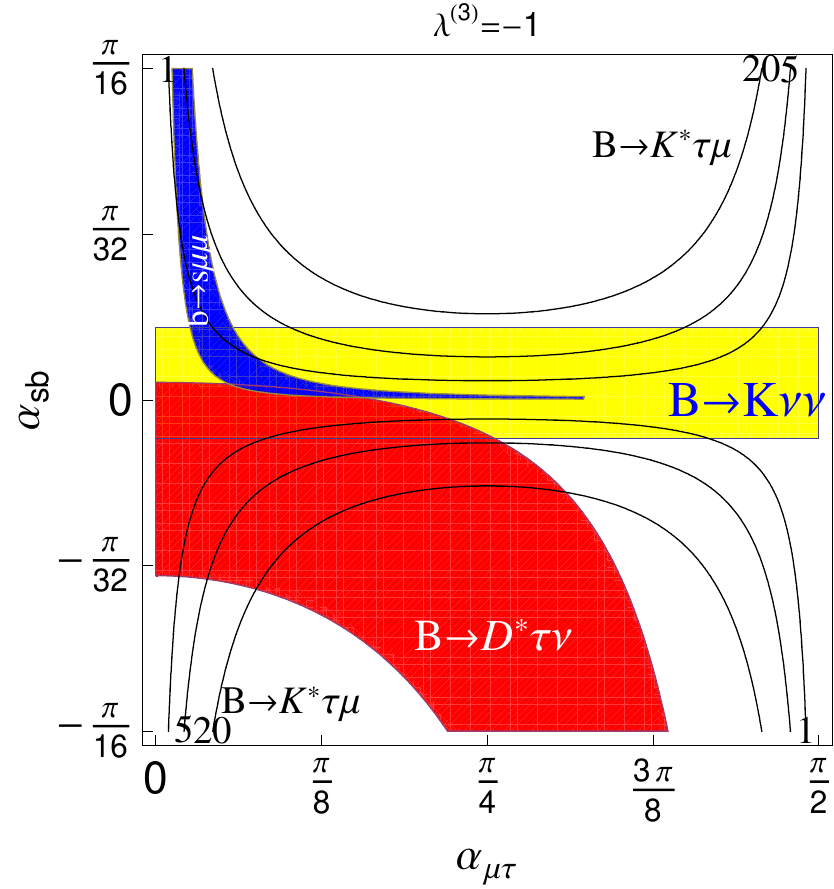}
\caption{Left: Allowed regions in the $m_{Z'}/g'$--$\sin (\theta_R)$ plane: the horizontal stripes correspond to $h\to\mu\tau$ ($1\sigma$) for $\tan\beta=85,\,50,\,25$ and $\cos (\alpha-\beta)=0.2$, (light) blue stands for (future) $\tau\to 3\mu$ limits at $90\%$~C.L. The gray regions are excluded by NTP or $B_s$--$\overline{B}_s$ mixing in combination with the $1\,\sigma$ range for $C_9$. $\theta_R$ is the tau-mu mixing angle. For details see Ref.~\cite{Crivellin:2015mga}.
Right: Allowed regions in the $\alpha_{\mu\tau}$--$\alpha_{sb}$ plane from $B\to K\nu\bar{\nu}$ (yellow), $R(D^*)$ (red) and $b\to s \mu^+\mu^-$ (blue) for $\Lambda=1\,$TeV and $\lambda^{(3)}=-0.5$ (left plot), $\lambda^{(3)}=-1$ (middle) and $\lambda^{(3)}=-2$ (right). Note that $\alpha_{sb}=\pi/64$ roughly corresponds to the angle needed to generate $V_{cb}$ and that if $\lambda^{(3)}$ is positive, $R(D^*)$ and $b\to s\mu^+\mu^-$ cannot be explained simultaneously. $\alpha_{\mu\tau}$ ($\alpha_{sb}$) is the misalignment angle between the 2nd and 3rd generation in the lepton (quark) sector. For details see Ref.~\cite{Calibbi:2015kma}.}
\label{fig:2}
\end{figure}

\begin{figure}[t]
\centering
\includegraphics[width=0.47\textwidth]{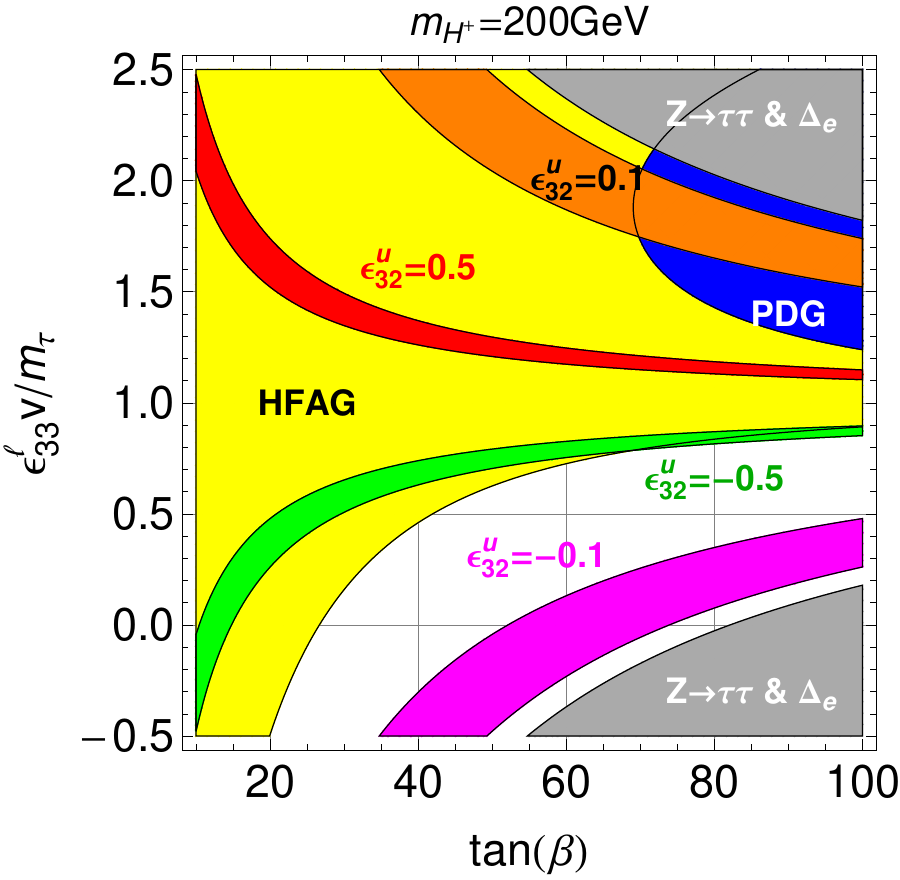}
\includegraphics[width=0.45\textwidth]{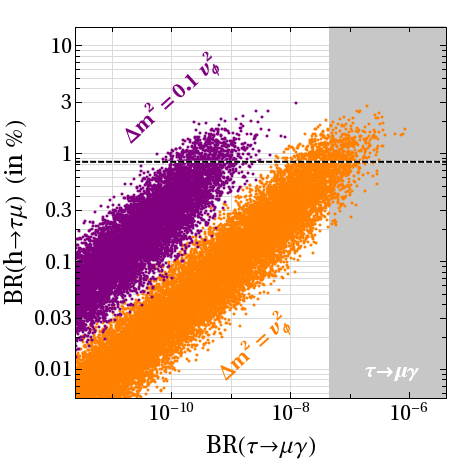}
\caption{Left: Allowed regions in the $\tan\beta$--$v/m_\tau\epsilon^\ell_{33}$ plane from $R(D^{(*)})$ and $\tau\to\mu\nu\nu$ at the $2\,\sigma$ level. The yellow region is allowed by $\tau\to\mu\nu\nu$ using the HFAG result for $m_H=30\,$GeV and $m_A=200\,$GeV, while the (darker) blue one is the allowed region using the PDG result. The red, orange, green, and magenta bands correspond to the allowed regions by $R(D^{(*)})$ for different values of $\epsilon^u_{32}$. The gray region is excluded by $Z\to \tau\tau$ and $\tau\to e \nu\nu$. For $m_H\simeq m_A$ the allowed regions from $\tau\to\mu\nu\nu$ would be slightly larger. For details see Ref.~\cite{Crivellin:2015hha}. 
Right: Correlations between $h\to\tau\mu$ and $\tau\to\mu\gamma$ for $\sin\alpha = 0.2$. The couplings $Y_{EL}, Y_{LE}, \lambda_{\mu L}, \lambda_{\mu E}, \lambda_{\tau L}, \lambda_{\tau E}$ (see Ref.~\cite{Altmannshofer:2016oaq} for the definitions) are scanned in the range $0.5 - 2$ and the masses of the vector-like leptons $M_E = M_L$ in the range $1~\text{TeV} - 3 ~\text{TeV}$ as well as $0.1 v < v_\phi < 2 v$. The gray region is excluded by the current bound on BR$(\tau\to \mu\gamma)$. The horizontal dashed line indicates the experimental central value of BR$(h\to\tau\mu)$.
\label{fig:3}}
\end{figure}

{\bf\boldmath $b\to s\ell^+\ell^-$:} Here a flavour changing neutral current is required which can be naturally generated at tree-level by a $Z'$ vector bosons~\cite{Descotes-Genon:2013wba,Gauld:2013qba,Buras:2013qja,Buras:2013dea,Altmannshofer:2014cfa,Crivellin:2015mga,Crivellin:2015lwa,Niehoff:2015bfa,Sierra:2015fma,Crivellin:2015era,Celis:2015ara} or by leptoquarks~\cite{Gripaios:2014tna,Becirevic:2015asa,Varzielas:2015iva,Alonso:2015sja,Calibbi:2015kma,Barbieri:2015yvd}. However, also NP contributing via loopa are possible~\cite{Gripaios:2015gra,Bauer:2015knc}.

{\bf\boldmath $B\to D^{(*)}\tau\nu_\tau$:} Here a tree-level NP contribution is required in order to generate the desired effect of the order of 25\% compared to the SM. Charged Higgses~\cite{Crivellin:2012ye,Tanaka:2012nw,Celis:2012dk,Crivellin:2013wna,Crivellin:2015hha} are one possibility, leading to large effects in the $q^2$ distribution. In addition, leptoquarks provide a valid explanation~\cite{Fajfer:2012jt,Deshpande:2012rr,Sakaki:2013bfa,Alonso:2015sja,Calibbi:2015kma,Bauer:2015knc,Fajfer:2015ycq,Barbieri:2015yvd} but also charged vector bosons are possible~\cite{Greljo:2015mma}. 

{\bf\boldmath $a_\mu$:} NP in $b\to s\mu^+\mu^-$ should also contribute to the AMM of the muon. Explanations beside supersymmetry (see for example Ref.~\cite{Stockinger:2006zn} for a review) include leptoquarks~\cite{Chakraverty:2001yg,Cheung:2001ip,Bauer:2015knc}, new scalar contributions in two-Higgs-doublet models (2HDM)~\cite{Broggio:2014mna,Crivellin:2015hha,Batell:2016ove}, and very light $Z^\prime$ bosons~\cite{Langacker:2008yv,Heeck:2011wj}.

{\bf \boldmath $h\to\mu\tau$:} Since the $B$ physics anomalies are related to $\tau$ and $\mu$ leptons, a connection to $h\to\mu\tau$ seems plausible. As the central value for the $h\to\mu\tau$ branching ratio is large, loop effects are in general not sufficient to generate the desired effect~\cite{Dorsner:2015mja}. Furthermore, also adding only vector-like fermions is not sufficient as the bounds from $\tau\to 3\mu$ and $\tau\to \mu\gamma$ are too stringent~\cite{Falkowski:2013jya,Herrero-Garcia:2016uab}. Therefore, introducing additional scalars is the most popular option (see e.g.~\cite{Campos:2014zaa,Heeck:2014qea,Crivellin:2015mga,Dorsner:2015mja,Huitu:2016pwk,Altmannshofer:2016oaq,Herrero-Garcia:2016uab}). 

\begin{figure*}[t]
\centering
\includegraphics[width=0.7\textwidth]{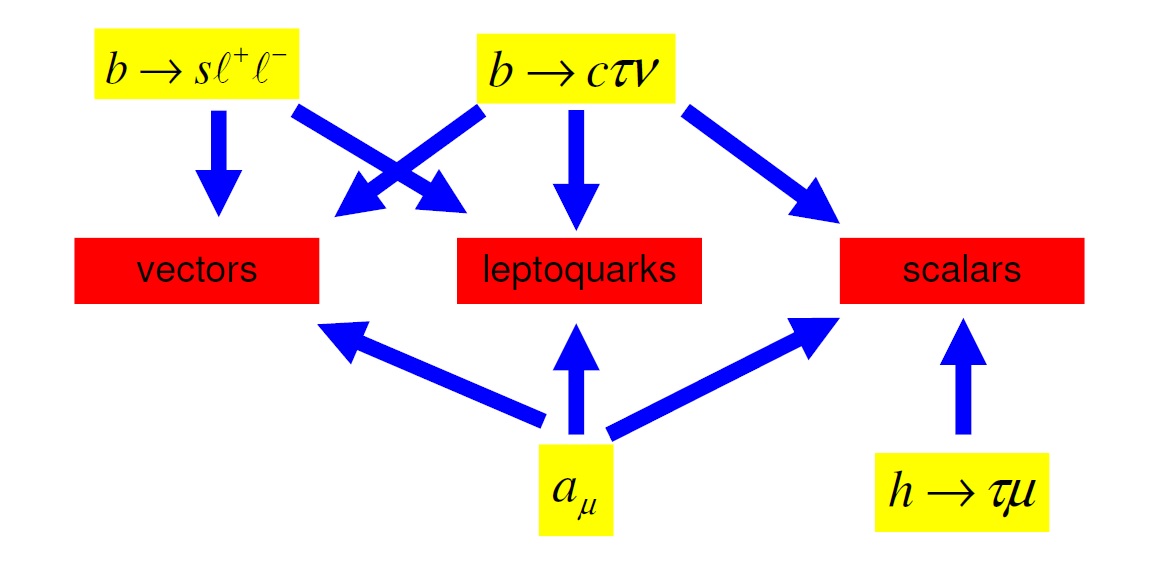}
\caption{Schematic picture of the implications for new particles from the various anomalies.\label{NPplot}}
\end{figure*}

\section{Selected models for simultaneous explanations of anomalies}

{\bf Multi Higgs {\boldmath  $L_\mu-L_\tau$} models: {\boldmath $h\to\tau\mu$} and {\boldmath $b\to s\mu^+\mu^-$} \cite{Crivellin:2015mga,Crivellin:2015lwa}}
\newline
Adding to a gauged $L_\mu-L_\tau$ model with vector like quarks~\cite{Altmannshofer:2014cfa} a second Higgs doublet with $L_\mu-L_\tau$ charge 2 can naturally give an effect in $h\to\tau\mu$ via a mixing among the neutral CP-even components of the scalar doublets~\cite{Heeck:2014qea}. In this setup a $Z'$ boson, which can explain the $b\to s\mu^+\mu^-$ anomalies, gives sizable effects in $\tau\to3\mu$ which are potentially observable at LHCb and especially at BELLE II (see left plot in Fig.~\ref{fig:2}). One can avoid the introduction of vector-like quarks by assigning horizontal charges to quarks as well~\cite{Crivellin:2015lwa}. In order not to violate the bounds from Kaon and $D$ mixing, the quarks of the first two generations must have the same charges. In this case, the effects in $b\to s$, $b\to d$ and $s\to d$ transitions are related in an MFV-like way by CKM elements, predicting an enhancement in $\Delta m_{B_s}$, $\Delta m_{B_d}$ and $\epsilon_K$ compared to the SM. Furthermore, as the $Z^\prime$ couples to light quarks, it has an observable cross section at the LHC.

{\bf \boldmath Leptoquarks: $b\to s\mu^+\mu^-$ and $b\to c\tau\nu$~\cite{Calibbi:2015kma}}
\newline 
While in $b\to c\tau\nu$ both leptoquarks as well as the SM contribute at tree-level, in $b\to s\mu^+\mu^-$ one compares a potential tree-level NP contribution to a loop effect\footnote{Alternatively, there is one leptoquark representation for which one can explain $b\to s\mu^+\mu^-$ by a loop effect and $b\to c\tau\nu$ at tree-level in the case on anarchic couplings~\cite{Bauer:2015knc} and even explain the AMM of the muon.}. However, $b\to c\tau\nu$ involves three times the third generation (assuming that the neutrino is of tau flavour in order to get interference with the SM contribution) but $b\to s\mu^+\mu^-$ only once. This suggests that leptoquarks with a hierarchical flavour structure, i.e. predominantly coupling to the third generation~\cite{Glashow:2014iga,Bhattacharya:2014wla}, can explain simultaneously $b\to s\mu^+\mu^-$ and $b\to c\tau\nu$ in case of a $C_9=-C_{10}$ (left-handed quark and lepton current) solution for $b\to s\mu^+\mu^-$. In this case one predicts sizable effects in $B\to K^{(*)}\tau\tau$, $B_s\to\tau^+\tau^-$ and $B_s\to\mu^+\mu^-$ below the SM, while the effects in $b\to s\tau\mu$ are at most of the order of $10^{-5}$ (see right plot in Fig.~\ref{fig:2}).

{\bf\boldmath 2HDM X: $a_\mu$ and $b\to c\tau\nu$~\cite{Crivellin:2015hha}}
\newline
In a 2HDM of type X, the couplings of the additional Higgses to charged leptons are enhanced by $\tan\beta$. As, unlike for the 2HDM II, this enhancement is not present for quarks, the direct LHC bounds on $H^0,A^0\to\tau^+\tau^-$ are not very stringent and also $b\to s\gamma$ poses quite weak constraints. Therefore, the additional Higgses can be light, which, together with the $\tan\beta$ enhanced couplings to muons, allows for an explanation of $a_\mu$. If one adds a coupling of the lepton-Higgs-doublet to third generation quarks ($\epsilon^u_{32}$), one can explain $b\to c\tau\nu$ as well by a charged Higgs exchange. In case of a simultaneous explanation of $a_\mu$ and $b\to c\tau\nu$ (without violating bounds from $\tau\to\mu\nu\nu$) within this model (see left plot of Fig.~\ref{fig:3}), sizable branching ratios (reaching even the \% level) for $t\to Hc$, with $m_H\approx 50-100\,$GeV and decaying mainly to $\tau\tau$, are predicted. Again, such a signature could be observed at the LHC.

{\bf\boldmath $L_\mu-L_\tau$ flavon model: $a_\mu$, $h\to\tau\mu$ and $b\to s\mu^+\mu^-$~\cite{Altmannshofer:2016oaq}}
\newline
In this model one adds vector-like leptons to the gauged $L_\mu-L_\tau$ model of Ref.~\cite{Altmannshofer:2014cfa}and one can explain $h\to\tau\mu$ via a mixing of the flavon (the scalar which breaks $L_\mu-L_\tau$) with the SM Higgs. Furthermore, one can account for $a_\mu$ by loops involving the flavon and vector-like leptons without violating the $\tau\to\mu\gamma$ bounds as this decay is protected by the $L_\mu-L_\tau$ symmetry (see right plot of Fig.~\ref{fig:3}). Despite the effects already present in the model of Ref.~\cite{Crivellin:2015mga}, one expects order one effects in $h\to\mu^+\mu^-$ detectable with the high luminosity LHC.

\section{Conclusions}
\label{conclusion}

In these proceedings we reviewed the anomalies in the flavour sector related to charged leptons together with some of their possible explanations. Interestingly, all anomalies involve muons and/or taus while the corresponding electron channels seem to agree with the SM predictions. This coherent picture of lepton flavour (universality) violation\footnote{For the implications in Kaon decays see~\cite{Crivellin:2016vjc}.} agrees with the stringent LEP constraints and suggests that the anomalies could be related, hinting at an unified explanation within a NP model. In Fig.~\ref{NPplot} we show in a schematic way which relations among the anomalies and new particles arise. Specific NP models can of course include the addition several new particles, potentially explain all anomalies and predict correlations among them and with other observables or processes detectable in future experiments.

\bibliography{BIB}


\end{document}